\documentclass[conference,10pt]{IEEEtran}
\IEEEoverridecommandlockouts

\usepackage[T1]{fontenc}
\usepackage{amsmath}
\usepackage{array}
\usepackage{booktabs}
\usepackage{cite}
\usepackage{graphicx}
\usepackage{tikz}
\usepackage{url}
\usepackage[hidelinks]{hyperref}
\usepackage{xspace}
\usetikzlibrary{arrows.meta,positioning}

\newcommand{\noema}{\textsc{Noema}\xspace}

\title{NOEMA: Executable Contracts for Learned Wireless Comparisons}

\author{
\IEEEauthorblockN{
Mostafa Naseri\IEEEauthorrefmark{1},
Mohamed Seif\IEEEauthorrefmark{2}\IEEEauthorrefmark{3},
H. Vincent Poor\IEEEauthorrefmark{3}, and
Adnan Shahid\IEEEauthorrefmark{4}}

\IEEEauthorblockA{
\IEEEauthorrefmark{1}Independent Researcher\\
\IEEEauthorrefmark{2}Department of Computer Science and Engineering,
Oakland University, Rochester, MI, USA\\
\IEEEauthorrefmark{3}Department of Electrical and Computer Engineering,
Princeton University, Princeton, NJ, USA\\
\IEEEauthorrefmark{4}IDLab, Department of Information Technology,
Ghent University--imec, Ghent, Belgium}

\thanks{The work of Adnan Shahid was supported in part by
the Horizon Europe program under the MSCA Staff Exchanges
2021 Grant Agreement No.~101086218 (EVOLVE project), and
in part by the Horizon-JU-SNS-2023 Research and Innovation
Program under Grant Agreement No.~101139194
(6G-XCEL project).}
}

\begin{document}
\maketitle

\begin{abstract}
Learning-based wireless research increasingly combines simulation, external model training, and benchmarking. Rebuilding these stages for every experiment is time-consuming, while differences introduced between them can silently change the conditions of a comparison. We present \noema, an open-source toolkit that prepares model
development and baseline evaluation from a shared, machine-checkable wireless scenario. From the same scenario, \noema can prepare benchmark execution, capture aligned training data, and export differentiable components for model development. Researchers can therefore focus on training and selecting their models, then return the selected component to the same controlled experiment for evaluation. \noema checks the returned model and verifies that declared comparison conditions, including available information, random conditions, and failure accounting, remain consistent through benchmarking and reporting. An included agentic-control example applies these contracts to a language-model supervisor that selects allocation policies and power budgets from historical link feedback. All 43 regression tests of the workflow checks produced their expected outcomes, including rejection of all 31 cases containing known inconsistencies. Separate fidelity tests showed that captured data and returned-model behavior were preserved. In a 20-replicate quadrature phase-shift keying phase-tracking
study, the mean bit-error-rate difference between the learned receiver and five-pilot smoothing was -0.01526
(95\% confidence interval: -0.01547 to -0.01506). \noema therefore provides both a reusable path for developing learned wireless components and a way to keep their final comparisons inspectable and checkable.
\end{abstract}


\section{Introduction}

Evaluating two receivers at the same SNR does not mean the comparison is matched. Differences in available information, such as pilot observations or channel state information, must also be declared and clarified.
Such  details in evaluation steps become especially important for learned wireless components.
In a communication system simulator, the simulator determines the channel, the impairments, and the random streams.
It can also capture the data that is visible to the learned component and the trainer (outside the simulator) trains the model.
A benchmark recipe imports that checkpoint alongside its baseline methods, learned or not, and the reporting stage selects and groups benchmark results for the final figures. Unrecorded changes to pilot visibility, signal normalization, or failure handling can invalidate the comparison. We present \noema\footnote{Source code and tutorials: \mbox{\href{https://github.com/M0574F4/noema-lab}{\texttt{github.com/M0574F4/noema-lab}}}.}, an experiment-control toolkit that prepares learning-based wireless experiments and
checks their declared comparison conditions. This is particularly important in fields where the researcher needs to compare the performance of their proposed component with several other baseline methods.

Sionna provides reusable differentiable components for
learning-based wireless experiments~\cite{hoydis2022sionna}.
Recent work explores both automating wireless
research~\cite{bjornson2026automating} and using agents to
observe and control radio networks~\cite{chatzistefanidis2025mxai}.
These directions motivate reusable experimental interfaces
with explicit information access, action limits, and
evaluation conditions.


\noema expresses the declared comparison conditions as
executable contracts: specifications that the workflow checks
at each supported boundary. A scenario describes the system's
operations, their inputs and outputs, and the experiment settings.
From this description, \noema prepares data capture, supported
differentiable exports, and benchmark execution. Differentiable
support components allow training gradients to pass through
fixed parts of the communication system, such as a simulated
channel, to the model being optimized.

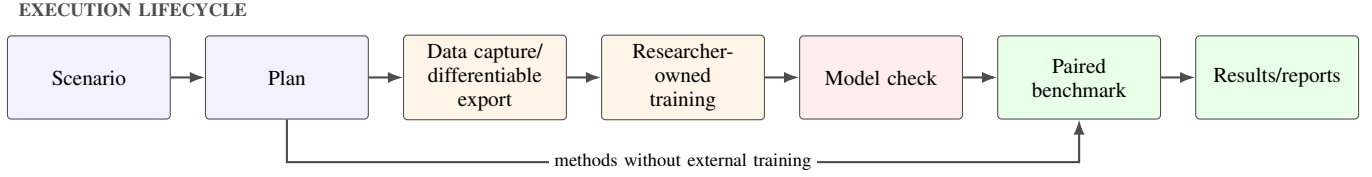
\begin{figure*}[t]
\centering
\begin{tikzpicture}[
  x=1cm,y=1cm,
  box/.style={draw=black!70,rounded corners=1.5pt,minimum height=1.10cm,
    text width=1.92cm,align=center,fill=blue!5,font=\footnotesize},
  learn/.style={box,fill=orange!9},
  gate/.style={box,fill=red!7},
  result/.style={box,fill=green!9},
  arrow/.style={-{Latex[length=2mm]},thick,draw=black!70},
  small/.style={font=\scriptsize,align=center}
]
\node[font=\scriptsize\bfseries,anchor=south west,text=black!70]
  at (-1.05,0.68) {EXECUTION LIFECYCLE};
\node[box] (scenario) at (0,0) {Scenario};
\node[box] (plan) at (2.62,0) {Plan};
\node[learn] (capture) at (5.24,0) {Data capture/\\differentiable export};
\node[learn] (train) at (7.86,0) {Researcher-owned\\training};
\node[gate] (return) at (10.48,0) {Model check};
\node[result] (bench) at (13.10,0) {Paired benchmark};
\node[result] (evidence) at (15.72,0) {Results/reports};
\draw[arrow] (scenario) -- (plan);
\draw[arrow] (plan) -- (capture);
\draw[arrow] (capture) -- (train);
\draw[arrow] (train) -- (return);
\draw[arrow] (return) -- (bench);
\draw[arrow] (bench) -- (evidence);
\draw[arrow]
  (plan.south) -- (2.62,-1.12) -- (13.10,-1.12) -- (bench.south);
\node[small,fill=white,inner xsep=2pt,inner ysep=0pt]
  at (7.86,-1.12) {methods without external training};
\end{tikzpicture}
\caption{\noema's lifecycle moves from scenario and planning, through data preparation and
researcher-owned training and selection, to returned-model checking, benchmarking, and reporting.
The lower path serves methods without external training.}
\label{fig:lifecycle}
\end{figure*}

Three mechanisms keep the workflow connected.
\emph{Plan continuity} checks that execution plans remain
consistent with the declared scenario.
\emph{Artifact admission} checks a returned model's requirements
and tensor interface before use.
\emph{Evidence closure} links reported measurements and figures
to the retained results.
Figure~\ref{fig:lifecycle} shows how these mechanisms fit around
researcher-controlled training and model selection.

Wireless simulators provide reusable physical-layer components
\cite{hoydis2022sionna,adler2022hermespy,mehlfuhrer2011vienna}, and learning-based methods can replace selected processing blocks \cite{oshea2017introduction}.
General-purpose resources such as OpenML benchmarking suites and Codabench support the organization and execution of comparable evaluations \cite{bischl2021openmlsuites,xu2022codabench}.
Workflow constraints and provenance records can also make assumptions checkable and document how results were produced \cite{schintke2024validityconstraints,leo2024workflowrunrocrate}.
\noema focuses specifically on the conditions that are important for benchmarking of learned wireless components.
It checks what information each block can access, and whether methods use random streams.
It also checks that resource use and failures are counted according to the declared rules for all compared methods.

This paper makes three contributions:

\begin{enumerate}
  \item An executable comparison contract that binds declared wireless conditions to scenario planning, learned-component, benchmark execution, and reporting.
  \item An open-source \noema prototype that derives benchmark,
dataset-capture, and differentiable-component export plans
from one scenario, while leaving model training and checkpoint
selection to the researcher.
  \item An evaluation of data fidelity, model admission, and comparison
  checks via a QPSK phase-tracking example.
\end{enumerate}

\section{Design: Keeping the Comparison Intact}

\subsection{What the contract binds}

Consider a phase-tracking task where the performance of a learned phase tracker is to be compared with 
a pilot smoother baseline method~\cite{mengali1997synchronization,khanzadi2011phasenoise}.
Comparison of these two methods depends on a number of factors including how the layout of each frame 
and pilot is organized, channel SNR, the type of carrier process the components need to correct, 
observations the method has access to, random streams used in the two experiments, and BER denominator.  
\noema represents these choices in a scenario directed acyclic graph (DAG) and its experiment-wide settings.
Table~\ref{tab:contract} connects each dimension of \noema to the execution evidence used to validate it.

Each operation in \noema declares requirements regarding inputs, outputs, parameters, what the data is 
(e.g. payload, coded bits, etc.), tensors, backend for execution of the operation, and its differentiability. 
In \noema, a scenario and its implementation are defined through a YAML or JSON recipe and the planner. 
Note that one might use different backend implementations, called for the same declared operation.

\begin{table*}[t]
\caption{The comparison contract connects scientific conditions to execution evidence.}
\label{tab:contract}
\centering
\footnotesize
\setlength{\tabcolsep}{4.5pt}
\begin{tabular}{@{}>{\bfseries\raggedright\arraybackslash}p{0.19\textwidth}
>{\raggedright\arraybackslash}p{0.36\textwidth}
>{\raggedright\arraybackslash}p{0.37\textwidth}@{}}
\toprule
Dimension & Declared conditions & Checked or retained evidence \\
\midrule
Graph and operations & DAG, typed ports, parameters, defaults, runner/backend capabilities
& Effective recipe, operation and implementation identities, frozen-plan identity \\
\addlinespace[2.5pt]
Population and information & Dataset version/split, preprocessing, capture taps, node-visible
information and CSI, model inputs & Source and split lineage, ordered capture records,
tensor-interface receipts \\
\addlinespace[2.5pt]
Randomness and signal & Master and per-operation streams, pairing policy, channel/impairments,
normalization target & Realized seeds/substreams, common-condition identity, per-source power
evidence \\
\addlinespace[2.5pt]
Resources and outcomes & Channel-use, time, energy, or memory budgets; outage and decode-failure
policy; denominator & Admission decision, attempted-item rows, failures, resource observations \\
\addlinespace[2.5pt]
Learned artifact & Replaceable role, checkpoint/component bytes, runtime, entrypoints, dtypes,
shapes, training-evidence locator & Package/component digests, inspected ONNX graph and interface,
predispatch admission \\
\addlinespace[2.5pt]
Measurement and report & Metric producer, unit, grouping, method order, required projections
& Metric provenance, result-local artifacts, data-bound tables/plots, verification checks \\
\bottomrule
\end{tabular}
\end{table*}

\subsection{Plan identities and boundary checks}


Let $G=(V,E)$ be the scenario DAG where $V$ and $E$ are components and connections between them, respectively. $K_v$ is the contract of operation $v$ and $r$ an execution purpose.
$M_{v,r}$ is the selected materialization for the component, $\theta_v$ its resolved parameters, and $X$ the settings that apply to the whole experiment.  The plan identity is
\begin{equation}
 I_r=H\!\left(G,r,X,\{K_v,M_{v,r},\theta_v\}_{v\in V}\right),
 \label{eq:plan}
\end{equation}
where $H$ is SHA-256 over canonical, versioned JSON. Unsupported ports, parameters,
runner/backend combinations, capture boundaries, or differentiability requirements stop planning
before run effects are created.

A plan identity, $I_r$, identifies an object, while a boundary check determines whether that object belongs in the declared comparison. At boundary $j$, let $B_j$ select the receipt fields that must agree, $R(\cdot)$
denote an object's receipt, and $C_j$ collect the other boundary predicates.  For object $a$ and its
expected parent or plan $p$, admission requires
\begin{equation}
 A_j(a,p)=C_j(a,p)\wedge
 [\pi_{B_j}(R(a))=\pi_{B_j}(R(p))].
 \label{eq:join}
\end{equation}
Here $\pi_{B_j}$ only keeps the fields that are required at that boundary for the check.  The checks apply at
planning, dispatch, artifact admission, result verification, and reporting.

In the phase-tracking example, the plan records the packet
and pilot layout, carrier impairments, SNR, and random-stream
settings. An intentional SNR change produces a new plan
identity, whereas attaching a result to an analysis with
incompatible declared conditions fails the boundary check.
The tests in Section~\ref{sec:boundary} demonstrate both
outcomes. A selected tracker can be reused in a compatible
new plan without treating the new run as the same experiment.
Within a comparison, the receivers share the declared
payload and channel realizations, but may use different
feature transformations and model capacities.

\subsection{External Training and Model Return}

There are three ways \noema runs an experiment, namely \emph{benchmark runner}, \emph{dataset capture runner}, and \emph{differentiable export runner}. The \emph{benchmark runner} runs the experiment and records what happened and the produced result.  The \emph{dataset capture runner} runs the needed parts of the experiment and saves selected intermediate data. It also captures information such as split sizes, SNR values, and seed.
Finally, when training requires gradients through part of the communication system, the \emph{differentiable export runner} exports that supported part to a gradient-capable backend such as PyTorch or Sionna. Naturally non-differentiable operations such as hard CRC, packet failure, and byte parsing remain non-differentiable unless the user provides a gradient approximation. 

It is important to distinguish between runner and backend. The runner determines the execution purpose, while the backend determines the component implementation. \noema is not limited to one software framework and during benchmarking components can run using ONNX, compiled code, or separate programs, as long as that implementation was declared.
The tested implementation includes 23 scenario templates spanning 16 task identifiers across seven suites.
For supported scenarios, researchers can export a \emph{training
bundle} containing the model interfaces, data-access or capture
recipes, and any required differentiable support code. They then
supply and train a model and return it for comparison with the
provided baselines.

In the phase-tracking example, packet observations and
training-only phase targets support external training of
the TCN. The selected tracker is then returned for interface
checking and paired evaluation, as described in
Section~\ref{sec:case}.
The boundaries under control of \noema are clear. Decisions such as model design, training setup, stopping criteria, and checkpoint selection remain with the researcher. 
After training, the researcher returns a \emph{model package} containing the selected component and the metadata required for its admission.
If the file has changed, requires unsupported software, or its inputs/outputs do not match what the experiment expects, \noema refuses to use the learned component.

\noema also records what actually happens during the run, e.g. failed transmissions or other failed attempts. These records allow \noema to check the declared source data, shared random conditions, and signal-power assumptions.

The toolkit also includes a delayed-CSI allocation example
in which a language-model acts as a slow supervisor.
Between complete simulation runs, it uses permitted historical
feedback to select one of four allocation policies and one
of five power budgets. The contract fixes the observation and
action limits, while leaving the decision to the agent.
The harness records decisions, latency, failures, and fallbacks,
and supports evidence verification and replay of recorded
actions. Its reliability metric is predicted BLER, not
measured decoder failures.

\noema assumes its own execution environment and internal components are trustworthy. The external returned model is what it explicitly checks before use.
I.e., \noema  only verifies what it can observe at the boundaries between components.

\section{Evaluation of the Contract Boundaries}
\label{sec:boundary}

We evaluate whether \noema preserves captured data and
returned-model behavior, detects incompatible experimental
conditions, and supports evaluating a selected model in a
fresh paired benchmark.
This section evaluates the first two properties. It also measures the storage and time cost of packaging and verifying the experiment evidence.
Section~\ref{sec:case} answers the third question and evaluates the selected model against the baseline methods.

\begin{table*}[t]
\caption{Retained evidence at the data, component, contract, and release boundaries.}
\label{tab:system}
\centering
\footnotesize
\setlength{\tabcolsep}{4.5pt}
\begin{tabular}{@{}>{\bfseries\raggedright\arraybackslash}p{0.17\textwidth}
>{\raggedright\arraybackslash}p{0.34\textwidth}
>{\raggedright\arraybackslash}p{0.40\textwidth}@{}}
\toprule
Boundary & Exercise & Observed result \\
\midrule
Returned execution & Direct vs. admitted QPSK-demapper and DeepJSCC ONNX paths & QPSK LLR
maximum absolute error 0; 100\% hard-decision agreement.  DeepJSCC encoder/decoder maximum errors
$1.79\!\times\!10^{-7}$/$5.96\!\times\!10^{-8}$; quantized pixels and PSNR exact \\
Capture alignment & 32 QPSK-demapper source runs; two selected taps; manifest order & Order and all
source identities retained; maximum symbol error 0; target-bit mismatches 0 \\
Contract behavior & 43 named regression cases & 43 expected
outcomes; all 31 declared inconsistencies rejected; no harness failures \\
Integrity controls & Incompatible mutations, intentional condition changes, unchanged controls &
Five mutations rejected; three changes produced new identities; two unchanged controls accepted \\
Release exercise & Two wheel builds; clean CPython 3.12 install; one 37.88 MiB, 592-file evidence bundle
packaged and verified five times & Wheel byte-identical; CLI/template/lint/run/UI/native ONNX pass;
median 0.571 s packaging, 9.925 s verification, 10.501 s per-trial total \\
\bottomrule
\end{tabular}
\end{table*}

\subsection{Preserving model behavior}

Importing a trained model into \noema should preserve its outputs.
We tested this by comparing direct execution of exported ONNX
models with execution through \noema for a QPSK demapper and a
DeepJSCC encoder--decoder for wireless image
transmission~\cite{bourtsoulatze2019deepjscc}.
For the QPSK demapper, both paths produced identical LLRs and
bit decisions. For DeepJSCC, the maximum absolute differences
were $1.79\times10^{-7}$ for the encoder and
$5.96\times10^{-8}$ for the decoder.
The reconstructed pixels and PSNR values
were identical, Table~\ref{tab:system}.

\subsection{Preserving data}
We also tested whether the data captured for training matches the original simulation data. We compared a captured QPSK-demapper training dataset with  32 simulation runs from which it was generated, at two capture points. All samples remained linked to the correct source run and kept in the same order. The received symbols and target bits matched exactly.

\subsection{Distinguishing changes from inconsistencies}

In order to evaluate the ability of \noema to distinguish between changes and inconsistencies we designed 43 predefined regression tests. Of the 43 tests, 31 contained known inconsistencies, and \noema rejected all 31 as expected.
In addition to that, we performed 10 integrity tests in which different parts of the experiment were changed or left unchanged. \noema rejected 5 incompatible changes, assigned new identities to 3 valid changes, and accepted 2 unchanged cases.

\subsection{Packaging and evidence-verification measurements}

We built the \noema software package twice on the same Linux system. Both builds produced the same 1.96 MB package. We also installed \noema in a clean Python 3.12 environment without access to the source repository. The installed package passed the checks in Table~\ref{tab:system}.  The tested \emph{evidence bundle}, which collects the experiment records and results used for verification, contained 592 files and occupied 37.88 MiB.  Across five trials, creating the evidence bundle took a median of 0.571 s and verifying it took 9.925 s, for a total of about 10.5 s.  These measurements show only the time required to package and verify this evidence bundle. They do not measure the additional runtime introduced by \noema or the cost of each individual \noema stage.

\section{Paired QPSK Phase-Tracking Case Study}
\label{sec:case}

\subsection{Workload and receiver information}

In this paper we choose QPSK phase tracking as the case study because it contains several conditions that must remain consistent for a fair comparison.
In terms of parameters of the signal, each packet contains 1024 data bits, a 16-symbol preamble, and one pilot after every 16 data symbols.  Each packet therefore contains 512 data symbols and 48 known
symbols---16 preamble symbols and 32 inserted pilots---for 560 transmitted symbols in total.

Unit-power QPSK symbols pass through AWGN at the specified $E_s/N_0$. The channel also adds three carrier impairments: a random initial phase, a random frequency offset, and time-varying phase noise. It has an initial phase uniform on $[-\pi,\pi]$, carrier-frequency offset uniform on $[-0.01,0.01]$ cycles/symbol, and Wiener phase increments with standard deviation $0.04$ rad.  These structures follow standard
carrier recovery models \cite{mengali1997synchronization,khanzadi2011phasenoise}.

\begin{table}[t]
\caption{Receiver methods and information used at inference.
No method receives simulator phase truth.}
\label{tab:receivers}
\centering
\footnotesize
\setlength{\tabcolsep}{2.4pt}
\begin{tabular}{@{}
>{\raggedright\arraybackslash}p{0.28\columnwidth}
>{\raggedright\arraybackslash}p{0.16\columnwidth}
>{\raggedright\arraybackslash}p{0.50\columnwidth}@{}}
\toprule
Receiver & Processing & Inputs and settings \\
\midrule
Uncompensated & streaming & received I/Q \\
Pilot interpolation & buffered &
public pilots; linear interpolation \\
Five-pilot smoothing & buffered &
five nearest public pilots; local linear fits \\
Decision-directed PLL & causal &
preamble, pilots, hard decisions;
$\alpha=.12$, $\beta=.005$ \\
Learned residual TCN & buffered &
same smoother, raw I/Q, pilot innovation, fourth-power cue \\
\bottomrule
\end{tabular}
\end{table}

Table~\ref{tab:receivers} compares the learned receiver with
four other methods. The uncompensated receiver applies
no carrier correction. Pilot interpolation and five-pilot
smoothing estimate phase from known pilots, while the
decision-directed PLL updates its estimate causally using
pilots and detected symbols.

The learned temporal convolutional network (TCN) uses received
I/Q and pilot-derived features to predict a correction to the
five-pilot smoother. Simulator phase is used only for the
training loss, not at inference. The predeclared statistical
comparison is against five-pilot smoothing, whose output the
TCN corrects; the other receivers provide additional performance
comparisons. Because the TCN  uses data-symbol information
and has greater capacity, this comparison evaluates
complete receivers rather than isolating the effect of
learning alone.
\begin{figure*}[t]
\centering
\includegraphics[width=0.985\textwidth]{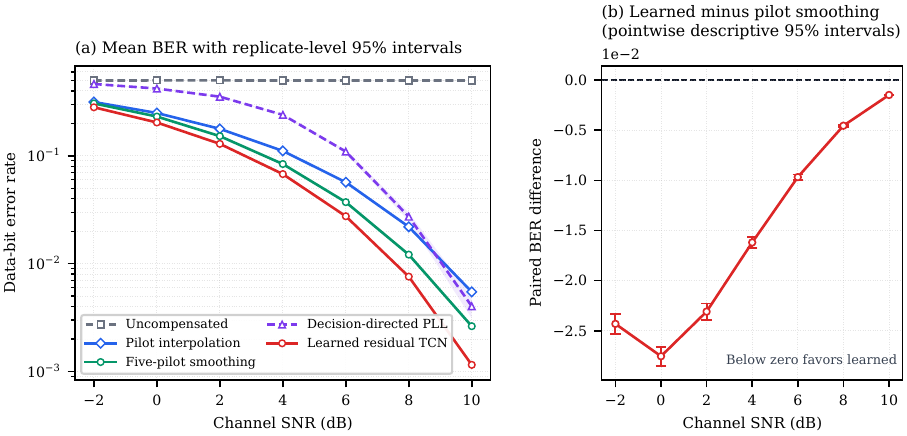}
\caption{Paired QPSK phase-tracking results:
(a) BER for five receiver-observable methods;
(b) learned-minus-five-pilot-smoothing difference.
Bands are pointwise descriptive 95\% intervals across
replicates; primary inference uses each replicate's average
difference over seven SNRs. The analysis repair is disclosed
in Section~\ref{sec:benchmark_protocol}.}
\label{fig:phase}
\end{figure*}
\subsection{Fixed component and fresh benchmark}
\label{sec:benchmark_protocol}
During development, we evaluated two TCN widths and two training seeds over 49 epochs. We selected the width-32 model from training seed 41 at epoch 47 and used that checkpoint unchanged in the final benchmark. We repackaged the selected ONNX model as a self-contained file.
To check that repackaging preserved its outputs, we ran the
original and repackaged versions on three deterministic input
cases separate from the final benchmark. Both versions received
identical inputs and produced bit-identical outputs in all
three cases.

Before generating or viewing the final benchmark results, we
fixed the five methods, 20 new random seeds, seven SNR values,
workload, primary baseline, performance measure, confidence
interval, and success criterion. The learned model had already
been selected. All 137 pre-execution checks passed, and the
benchmark random streams were separate from those used during
development.

At each replicate and SNR, all five methods use the same payload, noise, and carrier realization. Each method is evaluated on 256 packets, containing 262,144 data bits in total. The benchmark contains 20 replicates at seven SNR values, giving 140 paired comparison cases. With five methods per case, this produces 700 method executions and about 183.5 million bit decisions. A replicate is one complete repetition of the seven-SNR comparison using its own random seed. Each replicate
contributes one learned-minus-smoothing BER difference,
averaged over the seven SNRs. The confidence interval uses
these 20 averages; individual bits, method outputs, and
SNR points are not counted as additional replicates.

We evaluate each receiver at the seven channel-SNR values
$\Gamma=\{-2,0,2,4,6,8,10\}$ dB, where SNR denotes $E_s/N_0$.  For each replicate $s$, we compute the BER difference between the learned receiver and five-pilot-smoothing at each of the 7 SNR values and average over the 7 differences:
\begin{equation}
 d_s=\frac{1}{7}\sum_{\gamma\in\Gamma}
 \left(\mathrm{BER}_{L,s,\gamma}-\mathrm{BER}_{S,s,\gamma}\right),
 \label{eq:estimand}
\end{equation}
where $L$ is the learned method and $S$ is five-pilot smoothing.  We consider the comparison successful only if the 95\% Student's $t$ confidence interval for the mean learned minus smoothing BER difference, $\bar d\mathbin{\pm}t_{0.975,19}s_d/\sqrt{20}$, is negative, where $s_d$ is the sample standard deviation.  The interval describes
Monte Carlo variation under this fixed grid and simulated impairment; pointwise intervals in
Fig.~\ref{fig:phase} are descriptive.

The benchmark results were locked before analysis.
A later compatibility repair was needed to pass the 16
predeclared publication metrics from the larger stored results
to the analyzer. The repair was selected after outcomes had
become accessible and was not externally preregistered.
It removed no benchmark rows, changed neither the stored
results nor the statistical rule, and did not rerun the
benchmark. The operator reports not inspecting effect estimates
before choosing the repair; this cannot be independently verified.
\subsection{Paired result}

All 20 replicate results passed verification, including all 120 consistency checks. Across the 20 replicates, the learned receiver had an average BER that was $0.0152636$ lower than the five-pilot smoother. The variation across replicates was small (SD $0.0004306$, SE $0.00009628$).
The 95\% confidence interval for the BER difference was $[-0.0154652,-0.0150621]$.  Averaged equally across the seven SNR values, the learned receiver had a BER of $0.10269$ compared to $0.11795$ for five-pilot smoothing.  The learned receiver also had lower BER in each of the 140 replicate SNR comparisons. Figure~\ref{fig:phase}(a) shows lower mean BER for the learned
receiver than for the decision-directed PLL at every tested
SNR. The PLL improves sharply at higher SNR. A plausible
explanation is that incorrect detected symbols can disturb
its phase updates at low SNR, while more reliable decisions
improve its feedback at higher SNR. Pilot interpolation and
smoothing instead estimate phase from known pilots, while
the TCN uses buffered packet observations. The learned-versus-PLL
comparison therefore also reflects different causal and
buffered processing assumptions, rather than isolating the
benefit of learning.

Figure~\ref{fig:phase}(b) shows the mean paired difference
between the learned receiver and five-pilot smoothing at
each SNR. All seven differences are negative, favoring the
learned receiver. The largest absolute reduction occurs at
0 dB among the tested points; the gap narrows at higher SNR
as both BERs decrease. The pointwise intervals
describe variation across  20 replicates at each SNR,
whereas the primary confidence interval uses each replicate's
average difference over full SNR grid.

\section{Discussion and Related Work}

Existing tools already provide much of the infrastructure needed for wireless experiments. Sionna \cite{hoydis2022sionna} provides differentiable wireless components, HermesPy \cite{adler2022hermespy} supports simulation and hardware experiments, Vienna \cite{mehlfuhrer2011vienna} supports reproducible link evaluation, and Kaira provides PyTorch-based models and channels \cite{kaira2025}. Colosseum and POWDER provide facilities for hardware-based  experiments \cite{bonati2021colosseum,breen2021powder}, while SDP proposes a  protocol and benchmarking framework for wireless sensing \cite{zhang2026sdp}. \noema focuses on the comparison by checking that the conditions declared for the experiment remain consistent when a learned model is brought and compared with  baselines.
In this paper, we evaluate this workflow only in simulation.

General workflow systems already support portable execution \cite{crusoe2022cwl}, automatic checks of workflow assumptions \cite{schintke2024validityconstraints}, experiment tracking \cite{zaharia2018mlflow}, and records of artifacts and execution history \cite{soilandreyes2022rocrate,leo2024workflowrunrocrate}. These tools can also store experiment conditions, either directly or through extensions. \noema brings wireless-specific comparison checks together across training preparation, model return, execution, and reporting. It checks what information each method can use, whether methods are evaluated under the intended shared conditions, how resources and failures are counted, whether a  model is compatible, and whether reported results come from the retained benchmark. Our experiments test whether this \noema workflow behaves as intended. We do not compare it directly with a general workflow system configured to perform the same checks.

All experiments in this paper were designed and run by the authors on a single system running Ubuntu 24.04.4 LTS, equipped with an 11th Gen Intel Core i7-1185G7 processor (4 cores, 8 threads) and 16 GB of RAM. The phase-tracking experiment is our complete end-to-end case study and uses one impairment model. The other experiments test individual parts of \noema rather than a second full workflow. A broader evaluation should include other users, more channel conditions, different platforms, and hardware experiments. 
Development source is available at \url{https://github.com/M0574F4/noema-lab}. 

\section{Conclusion}

\noema turns a declared wireless scenario into a workflow that supports data capture, differentiable component export, benchmarking, and reporting while checking the conditions that make the comparison meaningful. This allows researchers to spend more of their effort on developing and training learned components, while \noema keeps the returned model connected to the experiment in which it is evaluated. Our evaluation shows that \noema can preserve the tested comparison conditions across model, benchmarking, and reporting, while keeping failed runs and later analysis changes visible. The main contribution is therefore a practical way to make learned-wireless experimentation easier to build and its comparisons explicit and checkable, rather than leaving those conditions spread across separate datasets, models, scripts, and figures.

\bibliographystyle{IEEEtran}
\bibliography{references}

@misc{hoydis2022sionna,
  author  = {Jakob Hoydis and Sebastian Cammerer and A{\"i}t Aoudia, Fay{\c{c}}al and Avinash Vem and Nikolaus Binder and Guillermo Marcus and Alexander Keller},
  title   = {Sionna: An Open-Source Library for Next-Generation Physical Layer Research},
  note = {arXiv preprint arXiv:2203.11854},
  year    = {2022},
  doi     = {10.48550/arXiv.2203.11854}
}

@article{adler2022hermespy,
  author  = {Jan Adler and Tobias Kronauer and Andr{\'e} Noll Barreto},
  title   = {{HermesPy}: An Open-Source Link-Level Evaluator for {6G}},
  journal = {IEEE Access},
  volume  = {10},
  pages   = {120256--120273},
  year    = {2022},
  doi     = {10.1109/ACCESS.2022.3222063}
}

@article{mehlfuhrer2011vienna,
  author  = {Christian Mehlf{\"u}hrer and Josep Colom Ikuno and Michal {\v{S}}imko and Stefan Schwarz and Martin Wrulich and Markus Rupp},
  title   = {The {Vienna} {LTE} Simulators: Enabling Reproducibility in Wireless Communications Research},
  journal = {EURASIP Journal on Advances in Signal Processing},
  volume  = {2011},
  number  = {1},
  pages   = {29},
  year    = {2011},
  doi     = {10.1186/1687-6180-2011-29}
}

@misc{kaira2025,
  author       = {{Kaira Contributors}},
  title        = {Kaira: A {PyTorch}-Based Toolkit for Simulating Communication Systems},
  year         = {2025},
  howpublished = {Software, version 0.2.1},
  url          = {https://github.com/ipc-lab/kaira/tree/v0.2.1}
}

@article{oshea2017introduction,
  author  = {Timothy J. O'Shea and Jakob Hoydis},
  title   = {An Introduction to Deep Learning for the Physical Layer},
  journal = {IEEE Transactions on Cognitive Communications and Networking},
  volume  = {3},
  number  = {4},
  pages   = {563--575},
  year    = {2017},
  doi     = {10.1109/TCCN.2017.2758370}
}

@book{mengali1997synchronization,
  author    = {Umberto Mengali and Aldo N. D'Andrea},
  title     = {Synchronization Techniques for Digital Receivers},
  publisher = {Plenum Press},
  address   = {New York, NY, USA},
  year      = {1997},
  doi       = {10.1007/978-1-4899-1807-9}
}

@inproceedings{khanzadi2011phasenoise,
  author    = {M. Reza Khanzadi and Hani Mehrpouyan and Erik Alpman and Tommy Svensson and Dan Kuylenstierna and Thomas Eriksson},
  title     = {On Models, Bounds, and Estimation Algorithms for Time-Varying Phase Noise},
  booktitle = {Proceedings of the 2011 5th International Conference on Signal Processing and Communication Systems},
  pages     = {1--8},
  year      = {2011},
  doi       = {10.1109/ICSPCS.2011.6140897}
}

@article{bourtsoulatze2019deepjscc,
  author  = {Eirina Bourtsoulatze and David Burth Kurka and Deniz G{\"u}nd{\"u}z},
  title   = {Deep Joint Source-Channel Coding for Wireless Image Transmission},
  journal = {IEEE Transactions on Cognitive Communications and Networking},
  volume  = {5},
  number  = {3},
  pages   = {567--579},
  year    = {2019},
  doi     = {10.1109/TCCN.2019.2919300}
}

@article{zaharia2018mlflow,
  author  = {Matei Zaharia and Andrew Chen and Aaron Davidson and Ali Ghodsi and Sue Ann Hong and Andy Konwinski and Siddharth Murching and Tomas Nykodym and Paul Ogilvie and Mani Parkhe and Fen Xie and Corey Zumar},
  title   = {Accelerating the Machine Learning Lifecycle with {MLflow}},
  journal = {IEEE Data Engineering Bulletin},
  volume  = {41},
  number  = {4},
  pages   = {39--45},
  year    = {2018}
}

@article{xu2022codabench,
  author  = {Zhen Xu and Sergio Escalera and Adrien Pav{\~a}o and Magali Richard and Wei-Wei Tu and Quanming Yao and Huan Zhao and Isabelle Guyon},
  title   = {Codabench: Flexible, Easy-to-Use, and Reproducible Meta-Benchmark Platform},
  journal = {Patterns},
  volume  = {3},
  number  = {7},
  pages   = {100543},
  year    = {2022},
  doi     = {10.1016/j.patter.2022.100543}
}

@inproceedings{bischl2021openmlsuites,
  author    = {Bernd Bischl and Giuseppe Casalicchio and Matthias Feurer and Pieter Gijsbers and Frank Hutter and Michel Lang and Rafael G. Mantovani and Jan N. van Rijn and Joaquin Vanschoren},
  title     = {{OpenML} Benchmarking Suites},
  booktitle = {Proceedings of the Neural Information Processing Systems Track on Datasets and Benchmarks},
  volume    = {1},
  year      = {2021}
}

@article{soilandreyes2022rocrate,
  author  = {Stian Soiland-Reyes and Peter Sefton and Merc{\`e} Crosas and Leyla Jael Castro and Frederik Coppens and Jos{\'e} M. Fern{\'a}ndez and Daniel Garijo and Bj{\"o}rn Gr{\"u}ning and Marco {La Rosa} and Simone Leo and Eoghan {{\'O} Carrag{\'a}in} and Marc Portier and Ana Trisovic and {RO-Crate Community} and Paul Groth and Carole Goble},
  title   = {Packaging Research Artefacts with {RO-Crate}},
  journal = {Data Science},
  volume  = {5},
  number  = {2},
  pages   = {97--138},
  year    = {2022},
  doi     = {10.3233/DS-210053}
}

@article{leo2024workflowrunrocrate,
  author  = {Simone Leo and Michael R. Crusoe and Laura Rodr{\'i}guez-Navas and Ra{\'u}l Sirvent and Alexander Kanitz and Paul {De Geest} and Rudolf Wittner and Luca Pireddu and Daniel Garijo and Jos{\'e} M. Fern{\'a}ndez and others},
  title   = {Recording Provenance of Workflow Runs with {RO-Crate}},
  journal = {PLOS ONE},
  volume  = {19},
  number  = {9},
  pages   = {e0309210},
  year    = {2024},
  doi     = {10.1371/journal.pone.0309210}
}

@article{zhang2026sdp,
  author  = {Di Zhang and Jiawei Huang and Yuanhao Cui and Xiaowen Cao and Tony Xiao Han and Xiaojun Jing and Christos Masouros},
  title   = {{SDP}: A Unified Protocol and Benchmarking Framework for Reproducible Wireless Sensing},
  journal = {arXiv preprint arXiv:2601.08463},
  year    = {2026},
  doi     = {10.48550/arXiv.2601.08463}
}

@article{crusoe2022cwl,
  author  = {Michael R. Crusoe and Sanne Abeln and Alexandru Iosup and Peter Amstutz and John Chilton and Neboj{\v{s}}a Tijani{\'c} and Herv{\'e} M{\'e}nager and Stian Soiland-Reyes and Bogdan Gavrilovi{\'c} and Carole Goble and {{The CWL Community}}},
  title   = {Methods Included: Standardizing Computational Reuse and Portability with the {Common Workflow Language}},
  journal = {Communications of the ACM},
  volume  = {65},
  number  = {6},
  pages   = {54--63},
  year    = {2022},
  doi     = {10.1145/3486897}
}

@article{schintke2024validityconstraints,
  author  = {Florian Schintke and Khalid Belhajjame and Ninon {de Mecquenem} and David Frantz and Vanessa Emanuela Guarino and Marcus Hilbrich and Fabian Lehmann and Paolo Missier and Rebecca Sattler and Jan Arne Sparka and Daniel T. Speckhard and Hermann Stolte and Anh Duc Vu and Ulf Leser},
  title   = {Validity Constraints for Data Analysis Workflows},
  journal = {Future Generation Computer Systems},
  volume  = {157},
  pages   = {82--97},
  year    = {2024},
  doi     = {10.1016/j.future.2024.03.037}
}

@misc{bjornson2026automating,
  author = {Emil Bj{\"o}rnson and Mischa Dohler and Jakob Hoydis and Heath, Jr., Robert W.},
  title = {Automating Wireless Research and Development: What Remains Human?},
  year = {2026},
  note = {Preprints.org preprint},
  doi = {10.20944/preprints202604.1172.v1}
}

@inproceedings{bonati2021colosseum,
  author    = {Leonardo Bonati and Pedram Johari and Michele Polese and Salvatore D'Oro and Subhramoy Mohanti and Miead Tehrani-Moayyed and Davide Villa and Shweta Shrivastava and Chinenye Tassie and Kurt Yoder and Ajeet Bagga and Paresh Patel and Ventz Petkov and Michael Seltser and Francesco Restuccia and Abhimanyu Gosain and Kaushik R. Chowdhury and Stefano Basagni and Tommaso Melodia},
  title     = {Colosseum: Large-Scale Wireless Experimentation Through Hardware-in-the-Loop Network Emulation},
  booktitle = {Proceedings of the 2021 IEEE International Symposium on Dynamic Spectrum Access Networks (DySPAN)},
  pages     = {105--113},
  year      = {2021},
  doi       = {10.1109/DySPAN53946.2021.9677430}
}

@misc{chatzistefanidis2025mxai,
  author = {Ilias Chatzistefanidis and Andrea Leone
    and Ali Yaghoubian and Mikel Irazabal and Sehad Nassim
    and Lina Bariah and Merouane Debbah and Navid Nikaein},
  title = {{MX-AI}: Agentic Observability and Control Platform
    for Open and {AI-RAN}},
  year = {2025},
  note = {arXiv preprint arXiv:2508.09197},
  doi = {10.48550/arXiv.2508.09197}
}

@article{breen2021powder,
  author = {Joe Breen and Andrew Buffmire and Jonathon Duerig
and Kevin Dutt and Eric Eide and Anneswa Ghosh and Mike Hibler
and David Johnson and Sneha Kumar Kasera and Earl Lewis
and Dustin Maas and Caleb Martin and Alex Orange and Neal Patwari
and Dan Reading and Robert Ricci and David Schurig and Leigh Stoller
and Allison Todd and Jacobus (Kobus) {Van der Merwe}
and Naren Viswanathan and Kirk Webb and Gary Wong},
  title   = {{POWDER}: Platform for Open Wireless Data-Driven Experimental Research},
  journal = {Computer Networks},
  volume  = {197},
  pages   = {108281},
  year    = {2021},
  doi     = {10.1016/j.comnet.2021.108281}
}

\end{document}